\documentclass[twocolumn]{aastex631}
\usepackage{multirow}
\usepackage{subfigure}
\def\msun{$M_{\odot}$} 
\def\lsun{$L_{\odot}$}

\def\xmm{{\it XMM-Newton}}

\tabletypesize{\tiny}
\shortauthors{Lin et al.}
\begin{document}

\title{Follow-up Observations of the Prolonged, super-Eddington, Tidal
  Disruption Event Candidate 3XMM~J150052.0+015452: the Slow Decline Continues}
\author{Dacheng Lin} 
\affiliation{Department of Physics, Northeastern University, Boston,
  MA 02115-5000, USA; d.lin@northeastern.edu}
\affiliation{Space Science Center, University of New Hampshire,
  Durham, NH 03824, USA}
\author{Olivier Godet}
\affiliation{IRAP, Universit\'{e} de Toulouse, CNRS, CNES, 9 avenue du Colonel Roche, 31028 Toulouse, France}
  \author{Natalie A. Webb}
\affiliation{IRAP, Universit\'{e} de Toulouse, CNRS, CNES, 9 avenue du Colonel Roche, 31028 Toulouse, France}
  \author{Didier Barret}
\affiliation{IRAP, Universit\'{e} de Toulouse, CNRS, CNES, 9 avenue du Colonel Roche, 31028 Toulouse, France}
  \author{Jimmy A. Irwin}
  \affiliation{Department of Physics and Astronomy, University of
    Alabama, Box 870324, Tuscaloosa, AL 35487, USA}
\author{S. Komossa}
 \affiliation{Max-Planck-Institut f\"ur Radioastronomie, Auf dem H\"ugel 69, 53121 Bonn, Germany}
  \author{Enrico Ramirez-Ruiz}
  \affiliation{DARK, Niels Bohr Institute, University of Copenhagen,
    Jagtvej 128, DK-2200, Copenhagen, Denmark}
  \affiliation{Department of Astronomy and Astrophysics, University of California, Santa Cruz, CA 95064, USA}
  \author{W. Peter Maksym}
  \affiliation{Center for Astrophysics \textbar\ Harvard  \&
    Smithsonian, 60 Garden St., Cambridge, MA 02138, USA}
  \author{Dirk Grupe}
  \affiliation{Space Science Center, Morehead State University, 235
    Martindale Drive, Morehead, KY 40351, USA}
  \author{Eleazar R. Carrasco}
  \affiliation{Gemini Observatory/NSF's NOIRLab, Casilla 603, La Serena, Chile}

\begin{abstract}
  The X-ray source 3XMM~J150052.0+015452 was discovered as a
  spectacular tidal disruption event candidate during a prolonged ($>11$
  yrs) outburst (Lin et al. 2017). It exhibited unique quasi-soft
  X-ray spectra of characteristic temperature $kT\sim0.3$ keV for
  several years at the peak, but in a recent \emph{Chandra}
  observation (10 yrs into the outburst) a super-soft X-ray spectrum
  of $kT\sim0.15$ keV was detected. Such dramatic spectral softening
  could signal the transition from the super-Eddington to thermal
  state or the temporary presence of a warm absorber. Here we report on
  our study of four new \emph{XMM-Newton} follow-up observations of
  the source. We found that they all showed super-soft spectra,
  suggesting that the source had remained super-soft for $>5$
  yrs. Then its spectral change is best explained as due to the
  super-Eddington to thermal spectral state transition. The fits to
  the thermal state spectra suggested a smaller absorption toward the
  source than that obtained in Lin et al. (2017). This led us to
  update the modeling of the event as due to the disruption of a
  0.75 \msun\ star by a massive black hole of \emph{a few}$\times10^5$ \msun. We
  also obtained two \emph{HST} images in the F606W and F814W filters
  and found that the dwarf star-forming host galaxy can be resolved
  into a dominant disk and a smaller bulge. No central point source
  was clearly seen in either filter, ruling out strong optical emission
  associated with the X-ray activity.

\end{abstract}


\section{INTRODUCTION}
\label{sec:intro}
Stars in close encounters with massive black holes (BHs) in galactic
nuclei could be tidally disrupted and subsequently accreted, resulting
in giant multiwavelength flares that could last for months to years
\citep{re1988, ko2015, ge2021}. Around $\sim$100 such tidal
disruption events (TDEs) have been found since the {\it ROSAT} All-Sky
Survey in 1990 \citep{koba1999}. Increasingly more TDEs were
discovered in recent years thanks to the large surveys: e.g., the
All-Sky Automated Survey for Supernovae (ASAS-SN; Shappee et al. 2014,
Kochanek et al. 2017), the Zwicky Transient Facility
\citep{vageha2021} and the All-Sky X-ray Survey by \emph{SRG}/eROSITA
\citep{sagime2021}. Most known TDEs were discovered in optical due to
the dominance of the optical surveys. Surprisingly,
they tend to be bright in optical but weak in
X-rays. Only a small fraction ($\sim$30\%) of known TDEs are bright in
X-rays, and they tend to be weak in optical. These X-ray TDEs mostly have super-soft X-ray thermal spectra of
characteristic temperatures $\sim$0.07 keV, and in some of them, the
soft-to-hard X-ray spectral state
transitions have been observed \citep[e.g.,][]{kohasc2004, wepava2021}.

However, the very special TDE candidate 3XMM~J150052.0+015452
(XJ1500+0154 hereafter) stood out in showing a unique spectral
evolution that had never been seen before \citep[][Lin17
hereafter]{liguko2017}. It is a prolonged TDE candidate coincident
with the nucleus of a dwarf star-forming galaxy at $z=0.14542$
($D_L=689$ Mpc, Lin17). Since an apparent fast rise (within months)
around 2005, it has remained X-ray bright (\emph{a few}$\times10^{43}$
erg s$^{-1}$) with very slow decay. In the peak, the source had
quasi-soft X-ray spectra of characteristic temperature $kT\sim0.3$
keV, and this phase lasted $\sim6$--10 yrs. Interestingly, in a deep
\emph{Chandra} observation at 10 yrs into the outburst, the source
clearly showed a different type of spectrum, super-soft or
$kT\sim0.15$ keV. There were two most likely explanations for such a
dramatic spectral softening. One is a spectral state transition, from
the super-Eddington state in the peak to the thermal state in the
decay. This explanation is strongly supported by its similarity to a
transient ultraluminous X-ray source in M31 \citep{mimima2013}, which
showed a very similar spectral evolution.  An alternative explanation
for the spectral softening is the presence of a transient highly
blueshifted ($\sim0.36c$) warm absorber. In any case, both
explanations implied the presence of a very long super-Eddington phase
in the peak.

Given its coincidence with a galactic nucleus, a natural explanation
for the large outburst of the source is a TDE. Its very slow decay and
unique spectral evolution, however, requires a different way of
modeling than those for other TDEs. First, the super-Eddington effects
have to be taken into account. When the accretion rate reaches above
some level, the inner accretion disk begins to reach the local
Eddington limit \citep{lireho2009}, and the luminosity will not
linearly scale with the mass accretion rate any more, but instead
probably logarithmically (thus with a lower radiative efficiency than
a standard disk), due to the presence of photon trapping and outflows
in the inner disk \citep{ohmi2007,krpi2012,kimu2016}. Secondly, the
very long super-Eddington phase requires either disruption of a very
massive ($\sim$10 \msun) star, which is expected to occur very rarely,
or slow circularization. In the standard TDE theory, the stellar
debris streams are quickly circularized and accreted after falling back
to the periapsis. However, several numerical studies showed that the
debris streams might intersect each other and get accreted at a much
larger distance than predicted in the standard theory
\citep{ko1994,gura2015,pisvkr2015,shkrch2015,hastlo2016}. In this
case, a long viscous timescale $\tau_\mathrm{visc}$, thus very slow circularization, is
expected, resulting in a fainter but longer TDE. Incorporating the super-Eddington effects and slow
circularization of $\tau_\mathrm{visc}=3$ yrs, Lin17 was
able to construct a TDE model of disrupting a $2$ \msun\ star by a
$10^6$ \msun\ BH to explain the
overall evolution of the event.

Since Lin17, we have obtained four new \xmm\ observations to monitor
the X-ray flux and spectral evolution of XJ1500+0154, in order to
further understand the cause of different spectral states and the nature of the event. We have also
obtained two \emph{HST} images to investigate the properties of its
host galaxy. In this Letter we present the results of these new
observations. In Section~\ref{sec:reduction}, we describe the data
analysis. In Section~\ref{sec:res}, we present the results. The
conclusions and the discussion of the nature of the event are given in
Section~\ref{sec:conclusion}.

\begin{figure*} 
\centering
\includegraphics[width=5.0in]{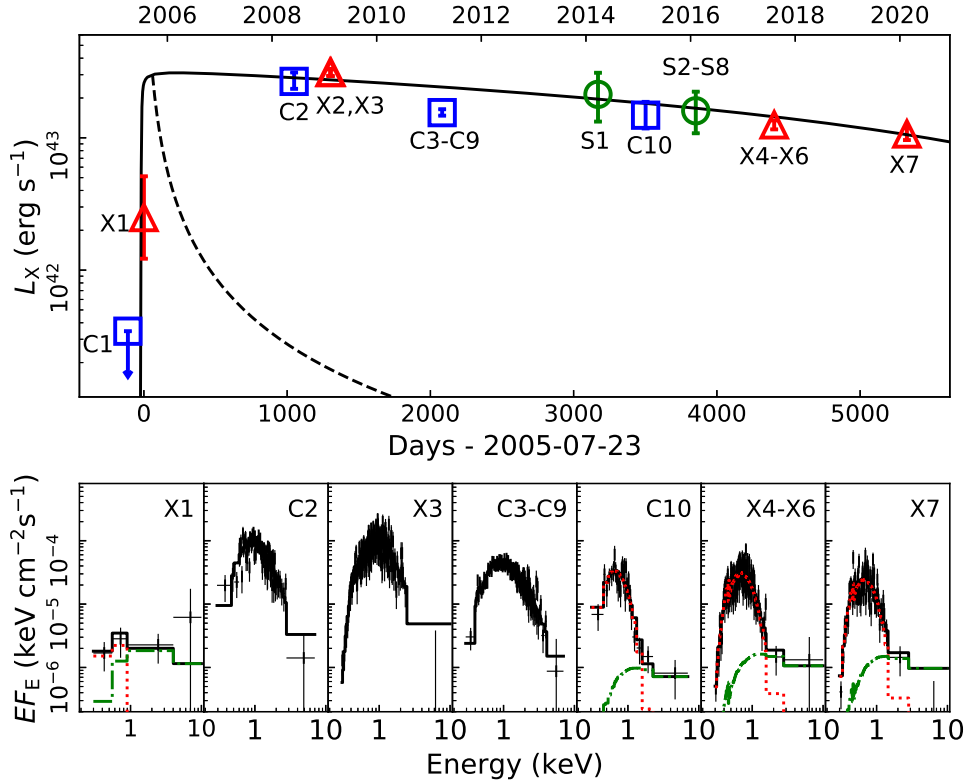}
\caption{Top panel: the long-term evolution of the X-ray luminosity
  (0.34--11.5 keV, source rest-frame, corrected for both Galactic and
  intrinsic absorption) from \xmm\ (red triangles), \emph{Chandra}
  (blue squares), and \emph{Swift} observations (green circles). Errors are at the
  90\% confidence level, except C1, for which the $3\sigma$ upper
  limit was given. The solid line is a model of disrupting a 0.75 \msun\ star by
  a BH of mass $2.2\times10^5$ \msun\ with slow circularization and
  super-Eddington effects (see text). The dashed line plots a standard TDE model
  of $t^{-5/3}$, assuming a peak X-ray luminosity the same as
  XJ1500+0154 and occuring at two months
  after the stellar disruption. 
 Bottom panels: sample unfolded X-ray spectra. X1, C10, X4--X6, and X7
 are fitted with a diskbb (red dotted line) plus PL (green
 dot-dashed line) model, while the others were fitted with the nthComp
 model. For visual purpose, the spectra are rebinned to
 be above $2\sigma$ in each bin in the plot, and for \emph{XMM-Newton}
 observations, only pn spectra are shown. \label{fig:ltlumsp}}
\end{figure*}

\begin{figure*} 
\centering
\includegraphics[width=3.2in]{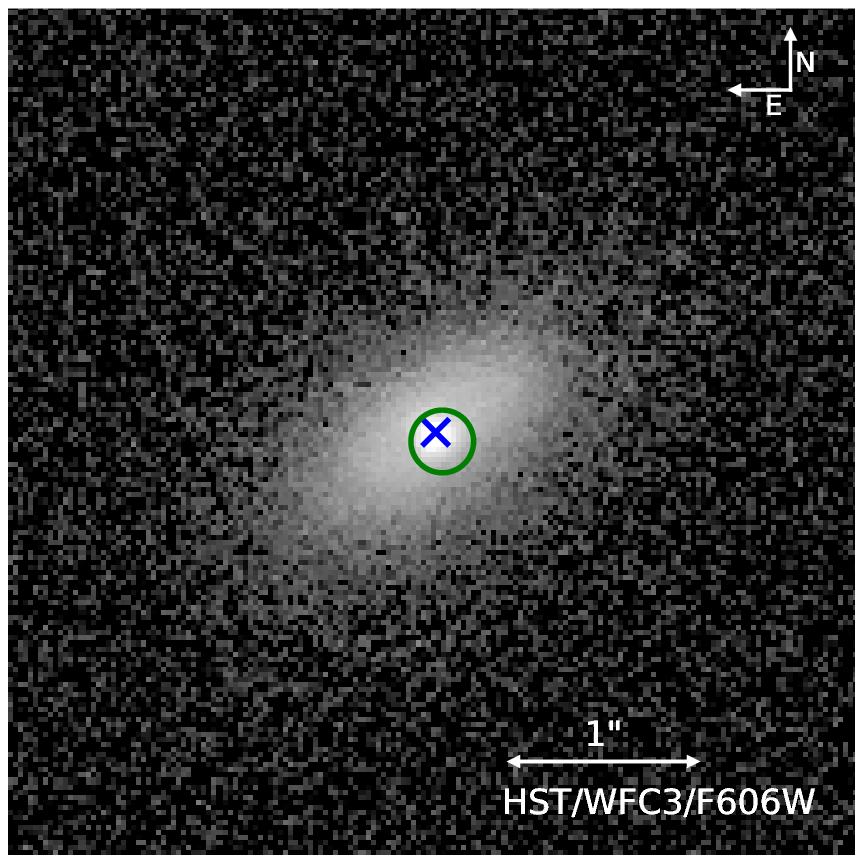}
\includegraphics[width=3.2in]{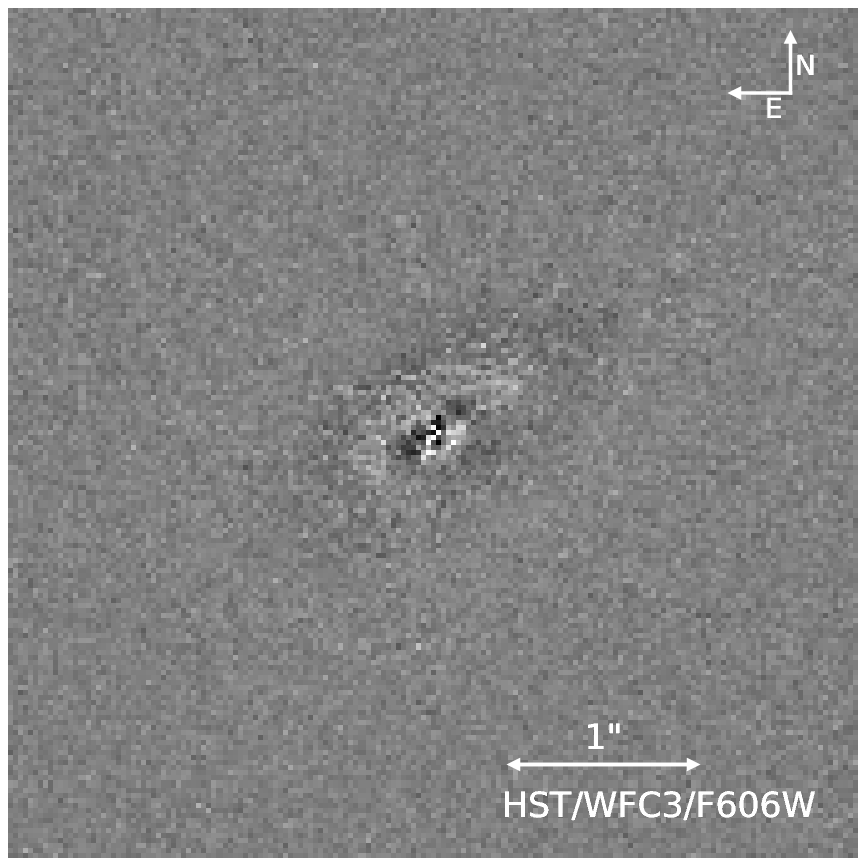}
\includegraphics[width=3.2in]{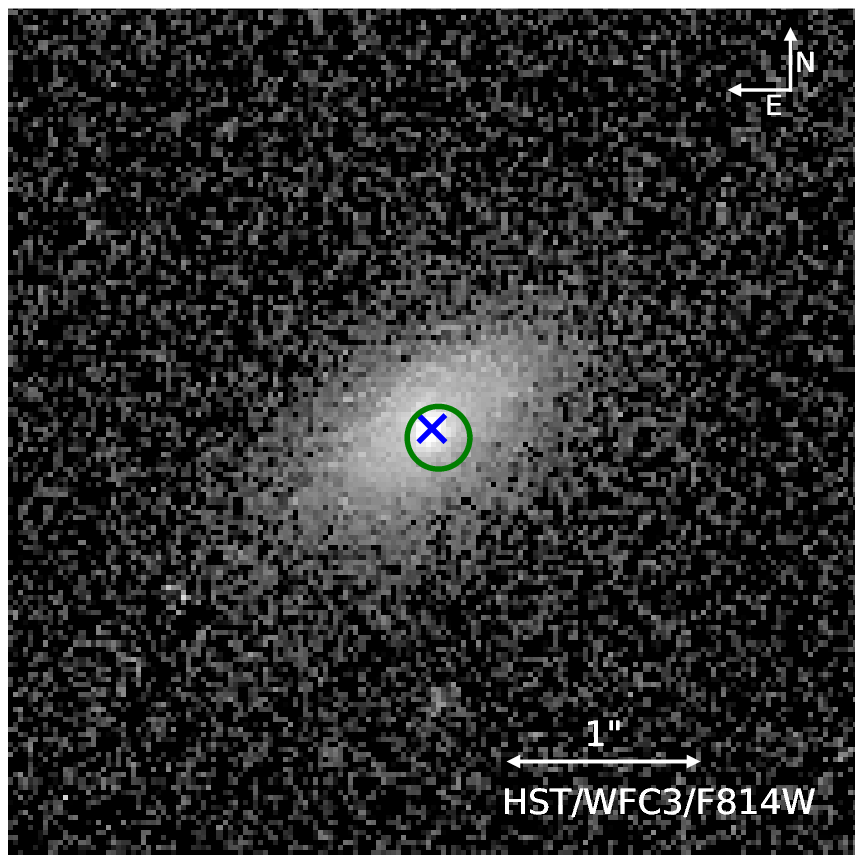}
\includegraphics[width=3.2in]{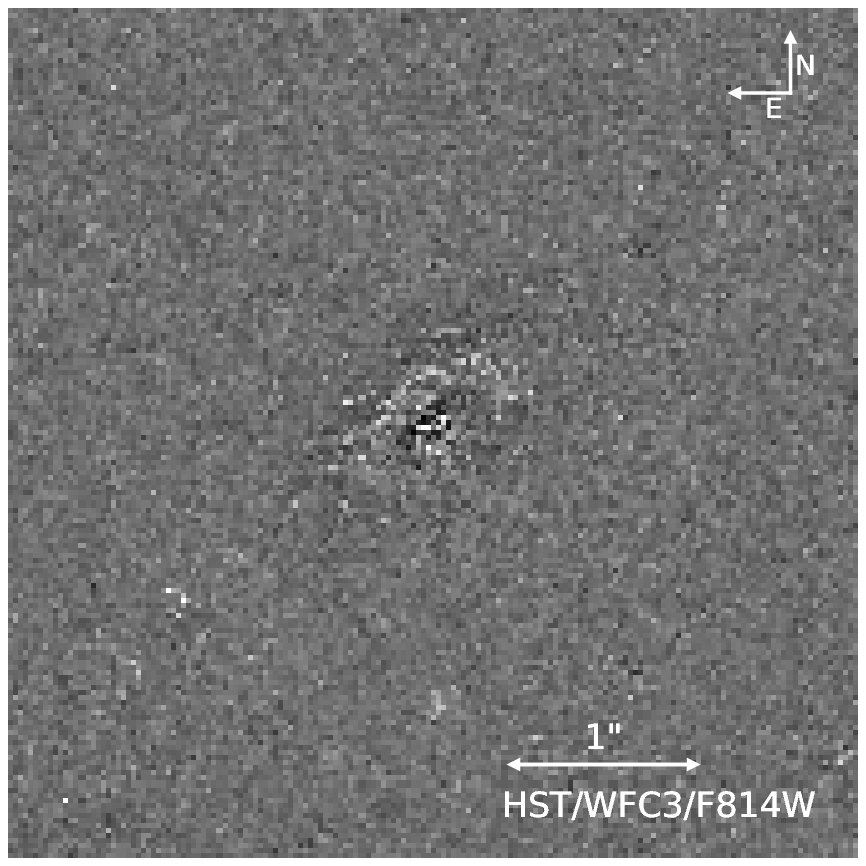}
\caption{Left panels: The two \emph{HST} images around the field of XJ1500+0154. The
  blue cross marks the center of its host galaxy, and the green
  circle of radius 0.18 arcsec (0.5 kpc) indicates the 95\% positional
  uncertainty of the source from the \emph{Chandra} observation
  C10. Right panels: GALFIT residuals with two S\'{e}rsic
  functions. 
  \label{fig:hstimg}}
\end{figure*}

\section{DATA ANALYSIS}
\label{sec:reduction}
There are seven \xmm\ observations (X1--X7 hereafter) of XJ1500+0154
in total thus far and three (X1--X3) were analyzed in Lin17. However,
for consistency, we reduced all \xmm\ observations again in the normal
way as we did for the first three observations in Lin17 but using SAS
18.0.0 and applying the updated calibration files as of 2020
October. In the new observations X4--X7, there were soft proton flares
from the Sun in X5--X7 in all cameras, most seriously in X5 and X6, but not in
X4. Data in bright background flares were excluded\footnote{https://www.cosmos.esa.int/web/xmm-newton/sas-thread-epic-filterbackground}. The source spectra
were extracted from a circular region of radius 20 arcsec and the
background spectra were extracted from a large nearby source-free region of
radius 50--100 arcsec. Table~\ref{tbl:obslog} lists the information of
all the \xmm\ observations. Due to the significant presence of the
high background flares, X5 and X6 have low statistics.  We combined
them with X4 to improve the statistics for spectral fits, because they
are close to X4 in time (within half a year) and turned out to have spectra
very similar to each other.

There were ten \emph{Chandra} (C1--C10 hereafter) and eight
\emph{Swift} (S1--S8 hereafter) observations. They had all been
analyzed by Lin17, and we used the same spectra obtained in that
study with one exception. In Lin17, S2--S5 and S6--S8 were shown to
have different spectra at the $2.5\sigma$ confidence level and were
combined separately into two spectra. Given that S2--S8 have very low
statistics and that they were taken very close in time (within two
weeks), we combined S2--S8 into a single spectrum. We note that the
\emph{Chandra} observations C3--C9, taken within two weeks, were also
combined into a single spectrum in Lin17, and it is also used in this
Letter.

The X-ray spectra from \xmm\ and \emph{Chandra} had enough counts
($>$ 5$\sigma$) and
were fitted within the X-ray fitting package XSPEC \citep[version
12.10.1,][]{ar1996}. Because the spectra generally do not have very
high statistics, we rebinned the source spectra to have at least one
count per bin and adopted the $W$ statistic, which is modified from
the $C$ statistic to account for the inclusion of background spectra \citep{waleke1979}. Similar to
Lin17, we applied the source redshift $z=0.14542$ to all the spectral
models with the convolution model \textit{zashift}. All models
included the Galactic absorption \citep[fixed at
$N_\mathrm{H}=4.4\times10^{20}$ cm$^{-2}$,][]{kabuha2005} using the
\textit{tbabs} model and the absorption intrinsic to the source using
the \textit{ztbabs} model. We used the \citet{wialmc2000} abundance
tables. We allowed the relative normalizations between \xmm\ cameras
to vary in order to  account for cross calibration uncertainties.

The two \emph{HST} images were taken on 2017 May 27 (about two months
before X4) under the program GO-14905. One image used the WFC3/F606W
filter (effective wavelength 5779 \AA) with four exposures of 375 s each (1500 s in total). The other
image was taken with the WFC3/F814W filter (effective wavelength 7968 \AA) and had two exposures of
351 s each (702 s in total).  The DrizzlePac software was used to
produce the drizzled stacked count images, one for each filter. The
pixel size was set to be 0.03 arcsec for both images. We fitted the
galaxy profile using the GALFIT software \citep{pehoim2010}. There are
not many bright stars for construction of the point-spread function
(PSF), and we used a nearby star to derive an empirical PSF.

\begin{table*}
\centering
\caption{The \emph{XMM-Newton} X-ray Observation
    Log. Columns: (1) the observation ID (our designation is given in
  parentheses), (2) the observation start date, (3) the smallest
  off-axis angle among all cameras, (4) the exposures of clean data after excluding
  periods of high background flares and (5) the 0.3--3 keV net source count
  rate (1$\sigma$ error) in the source extraction region for cameras pn, MOS1, and MOS2,
  respectively.}
\label{tbl:obslog}
\bigskip
\scriptsize
\sffamily
\begin{tabular}{rcccccc}
\hline
\hline
Obs. ID &Date &OAA &$T$ (ks) & Count rate (10$^{-3}$ counts~s$^{-1}$)\\
(1) & (2) &(3) & (4) & (5) \\
\hline
\multicolumn{4}{l}{\emph{XMM-Newton}:}\\
  \hline
 0302460101(X1) &2005-07-23 & 13\farcm5 & 22/33/33 &$1.23\pm0.46$/$0.78\pm0.24$/$0.24\pm0.20$\\
0554680201(X2) &2009-02-12 & 12\farcm5 & 43/-/64 &$37.3\pm1.0$/-/$11.3\pm0.5$\\
0554680301(X3) &2009-02-18 & 12\farcm5 & 42/-/64 &$35.1\pm1.0$/-/$9.6\pm0.4$\\
  0804370301(X4) & 2017-07-21 &   0\farcm17  & 15/18/18& $29.9\pm1.6$/$4.5\pm0.6$/$5.9\pm0.7$\\
   0804370401(X5) & 2017-08-09 &  0\farcm19 & 9/18/18& $25.9\pm1.9$/$4.2\pm0.6$/$4.9\pm0.6$\\
  0804370501(X6) & 2018-01-20 &   0\farcm17 & 5/15/15 &$33.1\pm2.9$/$6.0\pm0.8$/$5.6\pm0.7$\\
  0844040101 (X7) & 2020-02-21 &0\farcm16 & 22/29/29& $23.1\pm1.2$/$4.5\pm0.5$/$4.8\pm0.5$\\
\hline
\hline
\end{tabular}
\end{table*}

\begin{table*}
\centering
\caption{Spectral fit results for \textit{XMM-Newton} and \textit{Chandra} spectra. The
  fits used data within 0.3--10 keV for \textit{XMM-Newton} and data
  within 0.3--7 $\mathrm{keV}$ for \textit{Chandra}. All models include Galactic
  absorption of column density $N_\mathrm{H, Gal}=4.4\times10^{20}$
  $\mathrm{cm}^{-2}$ and the absorption intrinsic to the X-ray source at redshift
  0.14542 $N_\mathrm{H,
    i}$. The intrinsic absorption was fixed at $N_\mathrm{H,
    i}=2.65\times10^{21}$ $\mathrm{cm}^{-2}$,  the best-fitting value
  from the simultaneous fit to the C10, X4--X6, and X7 thermal-state spectra. $L_{\rm
    abs}$ and $L_{\rm unabs}$, in units of
  10$^{43}$ $\mathrm{erg}\ \mathrm{s}^{-1}$, are both the source rest-frame
  0.34--11.5 $\mathrm{keV}$ luminosity, with the former
  corrected for the Galactic absorption only and the latter corrected for both Galactic and intrinsic
  absorption. All errors given are at the 90\%-confidence
  level. Parameters without errors were fixed in the fits. The $W$ statistic values and
  the degree of freedown are given. }
\label{tbl:spfit}
\bigskip
\scriptsize
\sffamily
\begin{tabular}{r|c|c|c|cc}
\hline
\hline
Obs. & Model& Parameters  & $W (\nu)$& $L_{\rm abs}$ & $L_{\rm unabs}$\\

\hline
X1 &diskbb+PL  & $kT_{\rm diskbb}=0.11^{+ 0.08}_{-0.05}$ keV, $N_\mathrm{diskbb}=46^{+8095}_{-45}$, &197.4(192)& $ 0.07^{+ 0.04}_{-0.03}$ & $ 0.25^{+ 0.26}_{-0.13}$ \\
\ && $\Gamma_\mathrm{PL}=2.5$, $N_{\rm PL}=3.9^{+ 3.4}_{-2.9}\times10^{-6}$ & \\
C2 & nthComp & $kT_\mathrm{bb}=0.21\pm0.04$ keV, $kT_\mathrm{e}=1.0$ keV,  $\Gamma=4.25_{-0.51}^{+0.93}$ &$155.8(187)$ & $1.10\pm0.09$  & $2.67\pm0.39$ \\
  X2 & nthComp & $kT_\mathrm{bb}=0.23\pm0.03$ keV, $kT_\mathrm{e}=1.0$ keV,  $\Gamma=5.08_{-0.75}^{+1.65}$ &$579.7(593)$ & $ 1.26\pm0.06$  & $ 3.11\pm0.18$ \\
 X3 & nthComp & $kT_\mathrm{bb}=0.18\pm0.02$ keV, $kT_\mathrm{e}=1.0$ keV,  $\Gamma=4.02_{-0.36}^{+0.52}$ &$585.6(617)$ & $ 1.21\pm0.06$  & $ 3.11\pm0.20$ \\
  C3-C9 & nthComp & $kT_\mathrm{bb}=0.15\pm0.02$ keV, $kT_\mathrm{e}=1.0$ keV,  $\Gamma=3.32\pm0.13$ &$299.2(387)$ & $0.62\pm0.02$  & $ 1.55\pm0.09$ \\
C10 & diskbb+PL & $kT_{\rm diskbb}=0.148\pm0.013$ keV, $N_\mathrm{diskbb}=83.7^{+77.8}_{-40.3}$, & 67.4(62)&$ 0.34^{+ 0.06}_{-0.05}$  & $1.48^{+ 0.39}_{-0.31}$ \\
\ &&  $\Gamma_\mathrm{PL}=2.5$, $N_{\rm PL}=1.9^{+ 1.6}_{-1.1}\times10^{-6}$ &&\\
X4-X6 & diskbb+PL & $kT_{\rm diskbb}=0.167\pm0.009$ keV, $N_\mathrm{diskbb}=37.2^{+11.9}_{-9.0}$, & 517.7(563)& $ 0.33\pm0.02$ & $1.25\pm0.10$ \\
\ &&  $\Gamma_\mathrm{PL}=2.5$, $N_{\rm PL}=3.2\pm1.7\times10^{-6}$ &&\\
X7 & diskbb+PL & $kT_{\rm diskbb}=0.155\pm0.010$ keV, $N_\mathrm{diskbb}=45.8^{+21.2}_{-14.4}$, & 380.1(408)& $ 0.27\pm0.02$  & $1.07\pm0.11$ \\
\ &&  $\Gamma_\mathrm{PL}=2.5$, $N_{\rm PL}=2.9\pm1.9\times10^{-6}$ &&\\

\hline
\hline
\end{tabular}
\end{table*}

\begin{table*}
\centering
\caption{The profile fit results for the two \emph{HST} images with
  two S\'{e}rsic components using the software package GALFIT. All errors are at the 1$\sigma$ confidence level. For each
  S\'{e}rsic component, we list the AB magnitude, effective radius
  $r_\mathrm{e}$, index, axis ratio, and positional angle. The last
  column is the reduced $\chi^2$ value and the degrees of the freedom of
  the fit.}
\label{tbl:galfit}
\bigskip
\scriptsize
\sffamily
\begin{tabular}{r|ccccc|ccccc|c}
\hline
  \hline
 \ & &\multicolumn{3}{c}{Outer S\'{e}rsic} && &\multicolumn{3}{c}{Inner S\'{e}rsic} &&$\chi^2_\nu(\nu)$\\
Filter & AB mag & $r_\mathrm{e}$ (pc) & index & axis ratio & PA (deg) & AB mag
  & $r_\mathrm{e}$ (pc) & index & axis ratio & PA (deg)& \\
  \hline
  F606W & $20.83\pm0.01$ & $1415\pm6$ & $0.83\pm0.01$ & $0.50\pm0.01$& $-55.0\pm0.2$  & $22.28\pm0.01$ & $86\pm1$ &$1.77\pm0.07$ & $0.64\pm0.01$ & $-20.1\pm1.2$&1.19(39984)\\
  \hline
F814W & $20.20\pm0.01$ & $1428\pm12$ & $0.90\pm0.02$&$0.53\pm0.01$&$-54.9\pm0.5$&$21.83\pm0.02$&$81\pm2$&$1.47\pm0.15$&$0.75\pm0.02$&$-15.7\pm4.4$&1.16(39223)\\
  \hline
\end{tabular}
\end{table*}

\section{RESULTS}
\label{sec:res}

\subsection{X-ray Follow-ups}
The four new \xmm\ follow-up observations X4--X7 were to check whether
the X-ray spectra of XJ1500+0154 remained as super-soft as seen in the
last \emph{Chandra} observation C10 or were similar to the quasi-soft
spectra seen in the initial peak.  The fits to the new \xmm\ X-ray
spectra X4--X6 and X7 with the standard thermal-state model, a thermal
disk \citep[diskbb in XSPEC,][]{miinko1984} plus a weak powerlaw (PL), inferred a dominant
disk of $kT_\mathrm{diskbb}\sim0.15$ keV, thus similar to C10. This
means that the source has remained super-soft for at least $\sim5$
yrs. Then such spectra are very unlikely to be caused by transient
highly-blueshifted warm absorbers, which would require maintaining
such an extreme environment steadily for a long time. Instead, it seems
more natural to attribute them to the thermal state after transition
from the super-Eddington state in the peak. Hereafter, we will focus
on this scenario and explore the fits with the diskbb+PL model for
these spectra in detail.

In order to track the spectral evolution over time, it is desired to
tie and fix the intrinsic absorption column density $N_\mathrm{H,i}$
to the same value in the fits to all spectra. This assumes
$N_\mathrm{H,i}$ to be caused by the gas-rich host, not by the TDE itself.  In Lin17,
$N_\mathrm{H,i}$ was inferred by jointly fitting the quasi-soft
spectra, which occurred in the peak of the event and were most likely
due to Comptonization. The main problem for this method was that there
was a degeneracy between the seed photon temperature and the intrinsic
absorption, making the inference of $N_\mathrm{H,i}$ through the fits
to these spectra unreliable. Because the new \xmm\ observations strongly support that since C10 the source has
been in the well understood thermal state, which has a standard model
of a dominant thermal disk plus a weak PL, we changed to infer
$N_\mathrm{H,i}$ through the fits to the thermal state spectra
instead. After jointly fitting the three thermal-state spectra C10,
X4--X6 and X7 with $N_\mathrm{H,i}$ tied to be the same, we inferred
$N_\mathrm{H,i}=2.6\pm0.6\times10^{21}$ cm$^{-2}$. This intrinsic
column density was lower than that used in Lin17 ($4.3\times10^{21}$
cm$^{-2}$) and was adopted in all the final spectral fits that we will
present hereafter, unless specified otherwise.

For the quasi-soft X-ray spectra (C2, X2, X3 and C3--C9), Lin17 had
shown that the fits to these spectra with the thermal-state model
would infer unphysically high disk temperatures. Therefore Lin17
modeled them with the Comptonization model compTT. In this Letter,
besides adopting a new $N_\mathrm{H,i}$ value obtained above, we
changed to use the Comptonization model nthComp
\citep{zydosm1999,zdjoma1996,lizd1987} instead.  Although both nthComp
and compTT can provide a very similar quality of fits, the nthComp
model allows to specify disk-blackbody seed photons of the temperature
$kT_\mathrm{bb}$. The other parameters of the model are the electron
temperature $kT_\mathrm{e}$, asymptotic PL photon index $\Gamma$, and
normalization. Applying $N_\mathrm{H,i}$ obtained above, we find that
the electron temperature still cannot be constrained well in the fits
to individual spectra. Therefore we jointly fitted all the quasi-soft
X-ray spectra with their electron temperatures tied to be the same and
obtained $kT_\mathrm{e}=1.0_{-0.3}^{+2.5}$ keV. This value was fixed
in the final fits to the quasi-soft X-ray spectra.

The final spectral fit results are listed in Table~\ref{tbl:spfit} and shown in
Figure~\ref{fig:ltlumsp}, which plots the evolution of the unabsorbed
X-ray luminosity and X-ray spectra with time.  X4--X7 clearly show the
super-soft thermal state spectra as seen in C10. Although we observed
the overall decrease of the luminosity from C10 to X4--X7 by 20\%, the
inferred disk temperature is slightly higher in X4--X7
($kT_\mathrm{diskbb}\sim0.16$--0.17 keV) than in C10
($kT_\mathrm{diskbb}\sim0.15$ keV), at the $3\sigma$ confidence level. This could be due to a systematic
cross-calibration problem between different instruments. The disk
temperature was indeed lower in X7 than X4--X6 by 0.012 keV, but at
the confidence level of only $0.9\sigma$. Clearly further observation
over larger decay is needed to check whether the disk cools with
decreasing luminosity as expected for the thermal state.

For the quasi-soft X-ray spectra, the seed photons were inferred to
have temperature in the range of 0.15--0.23 keV, which is close to the
disk temperature obtained in the thermal state. This is a hot-seed
solution, while in Lin17, a cold-seed solution ($kT=0.04$ keV) was
adopted.

\emph{Swift} observations have too low statistics for meaningful
spectral fits, and their spectral states are difficult to
identify. The hardness ratio seems to support that S1 might have a
quasi-soft X-ray spectrum like C3--C9 (Lin17), and thus its luminosity
was inferred assuming the best-fitting model to C3--C9. S2--S8 were
one year after the first clear thermal state detection C10. Given that
the source had been consistently detected in the thermal state in the
four new \xmm\ observations after C10, we assumed that S2--S8 is also
in the thermal state and estimated its luminosity based on the fit to
C10.

Overall the unabsorbed X-ray luminosities are roughly a factor of 2.0
lower than those obtained in Lin17 due to a lower intrinsic absorption
inferred here. The new \xmm\ observations X4--X7 follow the global
trend of slow decay, from the unabsorbed rest-frame 0.34--11.5 keV
luminosity $1.5\pm0.3\times10^{43}$ erg s$^{-1}$ in C10 to
$1.3\pm0.1\times10^{43}$ erg s$^{-1}$ in X4--X6, and
$1.1\pm0.1\times10^{43}$ erg s$^{-1}$ in X7. These thermal-state
luminosities are lower than seen in the peak ($3.2\pm0.2\times10^{43}$
erg s$^{-1}$ in X2--X3).

\subsection{Host Galaxy Imaging}
The two \emph{HST} images of the host galaxy of XJ1500+0154 are shown
in the left panels in Figure~\ref{fig:hstimg}. This dwarf star-forming
galaxy seems to have a disk and a bulge, and the X-ray emission is
consistent with emanating from the center of the galaxy, based on the X-ray
position from C10. We therefore fitted the profile of the host galaxy
with two S\'{e}rsic functions, one for the disk and the other for the
bulge. The fitting residuals are shown in the right panels in Figure~\ref{fig:hstimg},
which indicate good fits. Table~\ref{tbl:galfit} lists the fitting
results. The disk component has an effective radius $\sim$1.4 kpc and
a S\'{e}rsic index $\sim$0.85 in both filters. The bulge component has
an effective radius $\sim$85 pc and a S\'{e}rsic index $\sim$1.6 in
both filters. The small S\'{e}rsic index of the bulge component is
consistent with the fact that this is a dwarf galaxy \citep{sagrma2013}. The
bulge is $\sim1.5$ mag fainter than the disk in both filters.

The measurement of the bulge luminosity can be used to infer the
central BH mass using the scaling relation between the BH mass and the
bulge. We estimated the $B$-band absolute magnitude of the bulge to be
$\sim-16.9$ mag (Galactic extinction corrected)
by linearly extrapolating the F606W and F814W photometry. Then based
on the
BH mass-bulge scaling relation from \citet{grsc2013},  we
inferred the BH mass to be \emph{a few}$\times10^5$ \msun, in
agreement with the value inferred from the X-ray spectral fitting (see
Section~\ref{sec:conclusion}).

The fit residuals in Figure~\ref{fig:hstimg} do not indicate the
presence of a central point source that can be associated with the
X-ray activity. We tried to add a point source in the fits. In
order to prevent it from converging to the fitting residuals
of the bulge, we forced the point source to reside right
at the center of the bulge in the fits. We found that the fits were hardly
improved with the addition of a central point source. For the F606W
image, the total $\chi^2$ value of the fit was just reduced by 17. The
inferred magnitude was $m_\mathrm{F606W}=24.3\pm0.3$ AB mag, or
$22.6$ AB mag after extinction correction (the extinction was
estimated assuming $\mathrm{E(B-V)}=1.7\times 10^{-22} N_\mathrm{H}$
and $N_\mathrm{H,i}=2.6\times10^{21}$ cm$^{-2}$ inferred from the
X-ray spectral fits). For the F814W image, the  total
$\chi^2$ value of the fit was not reduced at all by adding a central
point source.

 \section{DISCUSSION AND CONCLUSIONS}
 \label{sec:conclusion}
 The main result that we obtained from the four new \xmm\ follow-up
 X-ray observations of XJ1500+0154 is that the X-ray source seemed to
 have remained super-soft for at least $\sim$5 yrs since
 C10 on 2015 February 23, which strongly supports the identification of the thermal state
 for these observations. The new high-quality thermal state spectra can be fitted to
 infer the properties of the central BH with the more physical model optxagnf
 \citep{dodaji2012}. We jointly fitted the \xmm\ spectra X4--X6 and
 X7 (C10 was not included in order to minimize the cross-calibration systematic
 error, though the results turned out to be almost the same even if we included
 it). In order to model the thermal state spectra, we specified
in the model that the gravitational energy released
 in the disk is emitted as a color-corrected blackbody down to a
 coronal radius, within which the available energy is released in the
 form of a PL. The PL component was very weak, as in the fits with
 the diskbb+PL model. The PL index was fixed at 2.5. The free
 parameters included the Eddington ratio and the coronal radius. The
 BH mass, the BH spin and the intrinsic column density parameters were
 also left free but were tied to be the same for all the spectra. The inferred intrinsic column density $2.5\times10^{21}$
 cm$^{-2}$ is very close to that obtained with the diskbb+PL model. The
 BH mass and spin parameters are degenerate. We found that the
 dimensionless spin parameter $a_*$ needs to be $>0.8$, corresponding
 to a BH mass of $>2.2\times10^5$ \msun, if we required the thermal
 state luminosity to be sub-Eddington. When the maximal spin of
 $a_*=0.998$ was adopted, the maximal BH mass of $7.6\times10^5$
 \msun\  was obtained, with the luminosity corresponding to an Eddington
 ratio of 0.25.

Lin17 constructed a TDE model to explain the long-term evolution,
assuming disruption of a 2.0 \msun\ star by a non-spinning $10^6$
\msun\ BH. The model took into account slow circularization
($\tau_\mathrm{visc}=3$ yrs) and super-Eddington effects. To account
for the super-Eddington effects, the luminosity was assumed to scale
with the accretion rate until the mass accretion rate
$\dot{M}_\mathrm{d}$ reaches the 0.5 isotropic Eddington limit
$\dot{M}_\mathrm{Edd}$, above which, the luminosity scales with
$\dot{M}_\mathrm{d}$ as
$1.0+\log(\dot{M}_\mathrm{d}/0.5\dot{M}_\mathrm{Edd})$. The bolometric
correction factor from the X-ray unabsorbed luminosity was 4.0 in
Lin17. Here, we update the model, mostly driven by the use of a lower
intrinsic column density, which results in lower X-ray luminosities by
a factor of $\sim$2 and a lower bolometric correction factor (2.0
instead of 4.0).  Therefore the bolometric luminosities that we
inferred in this Letter are a factor of 4.0 lower than obtained in
Lin17. Applying the same model of the same $\tau_\mathrm{visc}$ value but assuming a $2.2\times10^5$ \msun\ BH of $a_*=0.8$, we find that the
disruption of a star of mass 0.75 \msun\ can describe the evolution of
the source very well (Figure~\ref{fig:ltlumsp}). Disruption of such a
much smaller star than obtained in Lin17 is more likely \citep[by a
factor of $\sim$3,][]{ko2016}. As noted in Lin17, there is a
degeneracy between the viscous timescale and the stellar mass, with a
more massive star needed for a smaller viscous timescale assumed.

Based on the new model, the total energy released and the total mass
accreted until X7 were
$2.0\times10^{52}$ ergs and 0.28 \msun, respectively. These values are
the highest among TDEs of small BHs \citep{liname2002,kohasc2004,vaanst2016,listro2020,mora2021}.

We obtained two \emph{HST} images of the host galaxy with the F606W
and F814W filters in 2017, when the X-ray source was bright in the
thermal state. The dwarf star-forming host galaxy was resolved into a
dominant disk and a smaller bulge. The images did not clearly show a
central point source that might be associated with the X-ray activity,
and we inferred $m_\mathrm{F606W}\gtrsim22.6$ AB mag (extinction
corrected). This corresponds to an absolute magnitude of
$\gtrsim-16.6$ AB mag and a luminosity
$\lambda L_\lambda<0.9\times10^{42}$ erg s$^{-1}$ or $<2.4\times10^8$
\lsun. This limit is a factor of 50 above the prediction from the fit
to the X4--X6 spectra using the optxagnf model, but it is low,
compared with other optical bright TDEs \citep[$\sim10^9$--$10^{10}$
$L_{\odot}$ at peak at the similar wavelength,
e.g.,][]{hokopr2016}. It is not clear whether this is because
XJ1500+0154 has a smaller BH than in optical bright TDEs or because it
is intrinsically weak in the optical like most X-ray TDEs. In
comparison, the peak absolute magnitude in F606W is about $-13.2$ AB
mag in the intermediate-mass BH TDE \object{3XMM J215022.4-055108}
\citep{listca2018}.

XJ1500+0154 remains one of the most spectacular TDE candidates thus
far. There have been some puzzling nuclear transients/outbursts of super-soft X-ray
spectra but with peculiar light curves
\citep[e.g.,][]{ligoho2017,misagi2019,rikalo2020,marame2021,armena2021},
and there are debates on whether they are TDEs or special active galactic
nuclei (AGN). It is hard to rule out the presence of the AGN optical signature
in XJ1500+0154 due to its modest absorption in the star-forming host
(Lin17). Its TDE nature can be tested with future long-term
X-ray monitorings. Our model predicts the luminosity to decay by one
order of magnitude from X7 in the next 14 yrs.

\begin{acknowledgments}

We thank the referee for the valuable comments that help improve the paper.  We would like to thank James
Guillochon for providing the updated modeling of the event. This work is supported by  the
National Aeronautics and Space Administration 
XMM-Newton GO program grants 80NSSC18K0718 and 80NSSC21K0619, by National Aeronautics and Space Administration
through grant number HST-GO-14905.001-A from the Space Telescope Science Institute, which is operated
by AURA, Inc., under NASA contract NAS 5-26555, and by the National Aeronautics and Space
Administration ADAP grant NNX17AJ57G. NW, OG and DB acknowledge CNES for financial support to the XMM-Newton Survey Science Center activities.
WPM acknowledges support from HST grants HST-GO-14272.011-A and
HST-GO-15351.001-A. Raw X-ray observations are available in the XMM-Newton Science Archive (\url{http://nxsa.esac.esa.int}); the Chandra data archive (\url{https://cxc.harvard.edu/cda/}, obsID:
5907, 9517, 12951, 12952, 12953, 13246, 13247, 13253, 13255, 17019),
and the NASA/{Swift} archive
(\url{http://heasarc.nasa.gov/docs/swift/archive}, obsID: 00033207001--00033207009).

\end{acknowledgments}

\software{SAS \citep[v18.0.0,][]{gadefy2004}, XSPEC \citep[v12.10.1,][]{ar1996}, GALFIT \citep{pehoim2010}, Astrodrizzle \citep{hadefr2012}}

\end{document}